\input{psfig}
\input{epsf}
\def\dofig#1{\vskip.2in\centerline{\epsfbox{#1}}}
\def\mev{\,{\rm Me\kern-0.1em V}}
\def\gev{\,{\rm Ge\kern-0.1em V}}

\documentstyle[12pt]{article}

\begin{document}
\vspace*{-1.25in}
\small{
\begin{flushright}
FERMILAB-PUB-97/121-T \\[-.1in] 
May, 1997 \\
\end{flushright}}
\vspace*{.75in}
\begin{center}
{\Large{\bf  Light Quarks, Zero Modes \\ 
and Exceptional Configurations}} \\
\vspace*{.45in}
{\large{W.~Bardeen$^1$, 
A.~Duncan$^2$, 
E.~Eichten$^1$, 
G.~Hockney$^1$ and
H.~Thacker$^4$}} \\ 
\vspace*{.15in}
$^1$Fermilab, P.O. Box 500, Batavia, IL 60510 \\
$^2$Dept. of Physics and Astronomy, Univ. of Pittsburgh, 
Pittsburgh, PA 15260\\
%$^3$insert George's address \\
$^4$Dept.of Physics, University of Virginia, Charlottesville, 
VA 22901
\end{center}
%%%%%%%%%%%%%%%%%%%%%%%%%%%%%%%
\vspace*{.3in}
\begin{abstract}
In continuum QCD, nontrivial gauge topologies give rise to 
zero eigenvalues
of the massless Dirac operator. In lattice QCD with Wilson 
fermions, these zero
modes appear as exactly real eigenvalues of the Wilson-Dirac 
operator and hence as poles in the 
quark propagator in the vicinity of critical hopping parameter. 
It is shown that ``exceptional configurations,'' which arise 
in the quenched approximation at small quark mass,
are the result of the fluctuation of the position of 
zero mode poles to sub-critical values of hopping parameter
on particular gauge configurations.  We describe a
procedure for correcting these lattice artifacts 
by first isolating the contribution of zero mode poles to 
the quark propagator and then shifting the sub-critical poles to the critical 
point. This procedure defines a modified 
quenched approximation in which accurate
calculations can be carried out for arbitrarily small quark masses. 
\end{abstract}

\newpage
\section{Introduction}

The study of QCD in the light quark limit is essential for 
understanding the chiral structure of pions and implications 
of the $U(1)$ anomaly which governs the generation of the  $\eta'$
mass as well as other applications involving the physics of light
hadrons.   In many previous studies using Wilson-Dirac 
fermions on the lattice, large statistical errors are encountered 
in calculations with light quarks and a small subset of the gauge 
configurations seem to play a dominant role in the final results.   
These ``exceptional configurations'' have prevented reliable 
studies of processes involving very light quarks.   
Instead, one has been forced to infer
the structure of light quark processes from extrapolations based on results 
obtained for quark masses substantially larger than the physical
masses of the up and down quarks.      In the quenched approximation, 
the large fluctuations resulting from exceptional
configurations appear to be related to the structure of 
the small eigenvalues of the Wilson-Dirac operator which can 
dominate the behavior of the quark propagators \cite{Mutter86}.
In this paper we show that 
this behavior is related to a specific artifact of the standard 
quenched lattice computations.   We introduce a 
modified quenched approximation (MQA) which removes the 
dominant artifact associated with the Wilson-Dirac formulation
of light fermions.   With this procedure, the 
problem of exceptional configurations is removed and the 
noise associated with light fermion computations is greatly 
reduced.   We also show that the same procedure must be 
applied to the standard $O(a)$ ``improved" actions, such as the 
Sheikholeslami-Wohlert (Clover) action\cite{SWaction},
and could play an important role in improving 
full unquenched QCD calculations.

In the next section we review the basics of the eigenvalue spectrum 
of the Wilson-Dirac operator and elucidate the relationship 
between a pole in a quark propagator at finite mass (sub-critical hopping
parameter $\kappa < \kappa_c$) and an exceptional gauge configuration. 
In Section 3 we show how to systematically remove these lattice
artifacts. This procedure defines a modified quenched approximation (MQA).
This MQA is then contrasted and compared with the unmodifed 
approach using physical observables in Section 4. In particular,
results for the pion mass versus quark mass are presented for
both Wilson and Clover actions.  The final section contains some
general remarks and possibilities for further study. 

\section{The Eigenvalue Spectrum and Zero Modes}

In the quenched approximation, the fermion determinant is
removed from the action functional during the Monte Carlo
generation of gauge field configurations, so that only the 
contributions of valence fermions are included.   This 
approximation makes the quenched formulation 
of the theory particularly sensitive to infrared structure 
involving light fermions.    This sensitivity depends on 
the particular formulation used to describe fermions in 
lattice simulations.   To understand this sensitivity, it is 
necessary to explore the structure of the fermion propagators.

The fermion propagator in a background gauge field, 
$A_{\mu}(x)$, may be written in terms of a sum over the eigenvalues 
of the Dirac operator,
\begin{equation}
{\cal D}f_i = \gamma*D f_i =\lambda_i f_i
\end{equation}    
and 	
\begin{equation}
S(x,y;\{A\}) = \sum_{eigenvalues} 
\frac{f_{i}(x;A)\bar{g}_{i}(y,A)}{(\lambda_i+m_0)}
\end{equation}       
where $f_i(x,A)$ and $\bar{g}_i(x,A)$ are the corresponding 
eigenfunctions.   The mass dependence of the propagator is 
determined by the nature of the eigenvalue spectrum of the 
Dirac operator.   In the continuum, the euclidean Dirac 
operator is skewhermitian and its eigenvalues 
are purely imaginary or zero.   Hence, the fermion 
propagators only have singularities when the real part of the 
mass parameter vanishes.    The behavior of the eigenvalue 
spectrum near zero determines the nature of dynamical 
chiral symmetry breaking \cite{Banks80}.   The zero eigenvalues, or zero 
modes, are related to topological fluctuations of the 
background gauge field by the index theorem associated with the 
chiral gauge anomaly \cite{tHooft,AStheorem,Smilga92}.    

The lattice formulation of Wilson-Dirac fermions 
qualitatively modifies the nature of the eigenvalue spectrum.   
The Wilson-Dirac operator is usually written as 
\begin{eqnarray}
\label{eq:Mdef}
& & {\cal D} \equiv D-rW \\
\label{eq:Ddef}
& & D_{a\alpha\vec{m},b\beta\vec{n}}=\frac{1}{2}(\gamma_{\mu})_{ab}
U_{\alpha\beta}(\vec{m}\mu)\delta_{\vec{n},\vec{m}+\hat{\mu}}-\frac{1}{2}(\gamma_{\mu})_{ab}U^{\dagger}_{\alpha\beta}
(\vec{n}\mu)\delta_{\vec{n},\vec{m}-\hat{\mu}} \\
\label{eq:Wdef}
& & W_{a\alpha\vec{m},b\beta\vec{n}}=\frac{1}{2}\delta_{ab}
(U_{\alpha\beta}(\vec{m}\mu)\delta_{\vec{n},\vec{m}+\hat{\mu}}
+U^{\dagger}_{\alpha\beta}(\vec{n}\mu)\delta_{\vec{n},\vec{m}-\hat{\mu}} )
\end{eqnarray}
where $U(\vec{n}\mu)$ are the link matrices associated with the lattice 
gauge fields, and the parameter $r$ is usually taken to be $1$.   
The Wilson-Dirac operator is neither skewhermitian (unless 
$r=0$) nor hermitian.   The Wilson term explicitly breaks chiral 
symmetry and lifts the doubling degeneracy of the pure lattice 
Dirac action.   As a result the eigenvalue spectrum of the 
Wilson-Dirac operator is no longer purely imaginary but fills 
a region of the complex plane. 
The discrete symmetries of the 
Wilson-Dirac operator imply that the eigenvalues appear in 
complex conjugate pairs, $(\lambda,{\lambda}^*)$ and obey 
reflection symmetries, $(\lambda,-\lambda)$.   In addition, 
there can be precisely real, nondegenerate eigenvalues
\cite{Smit87}.

\begin{figure}
\epsfxsize=14cm
\dofig{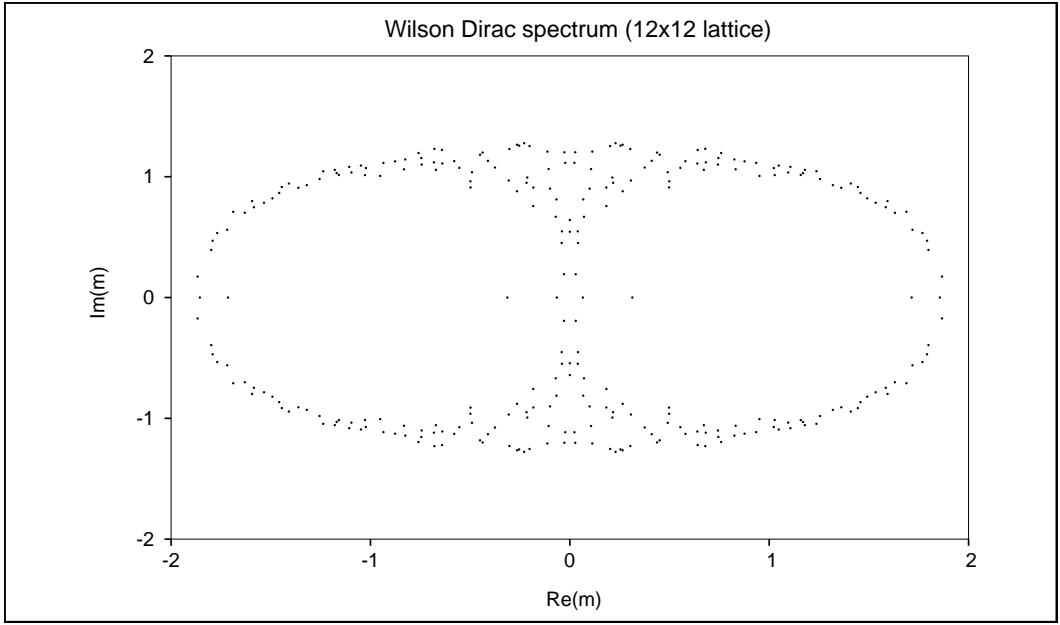}
\caption[]{Example of the eigenvalue spectrum for the Wilson-Dirac 
operator for QED2 on a $12 \times 12$ lattice with $\beta = 4.5$.}
\label{fig:eigen2d}
\end{figure}

Numerical studies of QED in two dimensions \cite{SmitVink87,QED2d97} 
and QCD in four dimensions \cite{Davies88,Verbaarschot96},
have confirmed the 
structure of the eigenvalue spectrum as well as the existence of 
isolated real eigenvalues.    In two dimensions, it is possible to study 
the spectrum of the Wilson-Dirac operator in great detail.   The
complex eigenvalue spectrum for a typical gauge configuration of 
QED2 on a $12\times 12$ lattice is shown in 
Figure \ref{fig:eigen2d}.   
%
%%% begin figure 1
%
%
%%% end figure 1
%
The left, right and central branches of the Wilson-Dirac
spectrum are clearly visible, as well as a pair of exactly 
real eigenvalues for each branch.   The left branch is usually 
associated with the spectrum of the continuum theory and the 
three other branches correspond to the spectrum of massive doubler modes 
for QED2; there would be fifteen doubler branches expected for 
a similar plot of QCD4.    The exactly real eigenvalues are the 
Wilson-Dirac analog of the continuum zero modes.   
Unlike the situation in the continuum, these real eigenvalues do not 
all occur at the specific value associated with the zero 
mass limit, even for real eigenvalues of a given gauge 
configuration.     Because of the fluctuations in the real 
part of the eigenvalue spectrum, the massless limit can only be 
defined through an ensemble average in Monte Carlo 
calculations.  We will see that the fluctuations in the position
of the zero modes are the primary reason for instabilities in 
calculations with light Wilson-Dirac fermions 
\cite{SmitVink87,Iwasaki89,Itoh87}.
These fluctuations are 
a direct result of the chiral symmetry breaking which occurs as an 
artifact of the Wilson-Dirac fermion formulation.    Shifts in the positions 
of the real eigenvalues will cause spurious poles in the 
fermion propagator for light fermion masses.    These nearby poles 
are the cause of the exceptional configurations 
encountered in quenched calculations with Wilson-Dirac 
fermions.

\begin{figure}
\epsfxsize=14cm
\dofig{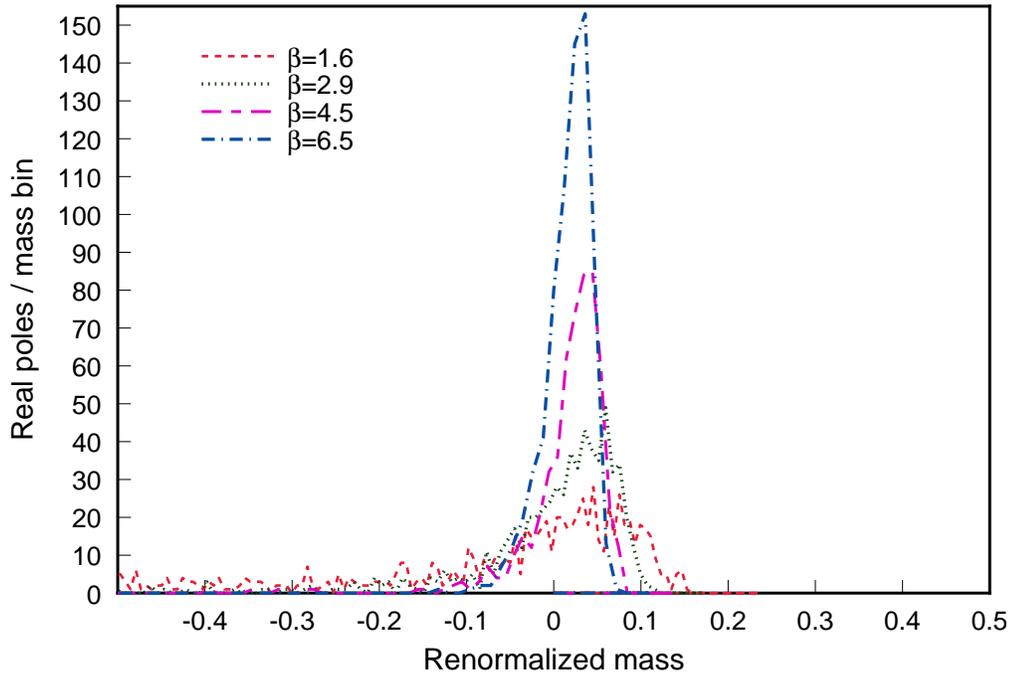}
\caption[]{Distribution of real eigenvalues in QED2 near 
zero fermion mass for $\beta = 1.6, 2.9, 4.5$ and $6.5$ on 
a $12\times 12$ lattice. Results shown here are for 
a sample of 1000 independent gauge configurations at each $\beta$. 
Fermion mass bins were 0.012 wide.} 
\label{fig:beta2d}
\end{figure}

The shifts in the real part of the Wilson-Dirac eigenvalue 
spectrum are lattice artifacts, as the non-skewhermitian part of the
lattice Wilson-Dirac operator is associated with higher dimension
 operators.    In the continuum limit, these artifacts
are expected to disappear.      In Figure \ref{fig:beta2d}, we show the real 
eigenvalue distribution for our QED2 lattice as a function of 
$\beta$.  
%
%%% begin figure 2
%
%
%%% end figure 2
%
For larger values of $\beta$, the distribution becomes 
more sharply peaked as expected for the approach to the 
continuum limit.    For any fixed $\beta$, the distribution 
maintains a finite width and the naive massless limit remains 
undefined.    Indeed the distribution of poles in the 
corresponding fermion propagators  leaves the naive 
quenched theory formally undefined.   For any fixed fermion 
mass the number of exceptional configurations associated with 
the zero mode poles can be expected to decrease as $\beta$ is 
increased, but the limit to infinite Monte Carlo statistics at 
fixed $\beta$ will not converge\cite{QED2d97}.

In four dimensions, it is more difficult to study the 
eigenvalue spectrum, even if we restrict our attention to just the
 small eigenvalues.
The zero modes appear as poles in the fermion propagator
\begin{eqnarray}
& & (D-rW)f_i = \lambda_i f_i \\
& & S(x,y;\{U(A)\})_{AB} = \sum_{eigenvalues}
\frac{f_{iA}(x;U)\bar{g}_{iB}(y,U)}{\lambda_i + 1/2\kappa} \\
& & m_{fermion} = 1/2\kappa - 1/2\kappa_c
\end{eqnarray}
where $\kappa$ is the hopping parameter, with the critical value $\kappa_c$
determined from the ensemble ensemble average pion mass.
For modes with $\lambda_i < -1/2{\kappa_c}$, the propagator 
has poles corresponding to positive mass values.   The 
position of these poles can be established by studying any 
smooth projection of the fermion propagator as a function of 
$\kappa$.    In our computations, we use the integrated 
pseudoscalar charge to probe for the shifted real eigenvalues
\begin{equation}
Q(\kappa) = \int d^4x<\bar{\psi}(x)\gamma_5\psi(x) >
\end{equation}
We employ the same method that was used by Kuramashi et al.\cite{Kuramashi94}
to study hairpin diagrams and the ${\eta}'$ mass.  A 
fermion propagator, $G(n,\kappa)$, is computed using a unit source at every 
site for each color and spin.  The result is traced over color 
and spin and summed over the lattice.     
\begin{equation}
Q(\kappa) \rightarrow {\rm Tr}(i\gamma_5 \sum_{n} G(n,\kappa))
\end{equation}
The color averaging over the lattice volume reproduces the 
integrated charge.   This is a global quantity which samples the 
full lattice volume.    By computing its value for a range of 
kappa values we can search for poles in the fermion 
propagator.    Near a pole,  we should find
\begin{equation}
Q(\kappa) \rightarrow \frac{R}{1/\kappa-1/\kappa_{pole}}
\end{equation}
In the continuum, the residue R would be directly proportional 
to the global winding number of the gauge field
configuration \cite{SmitVink87}.

\begin{figure}
\epsfxsize=14cm
\dofig{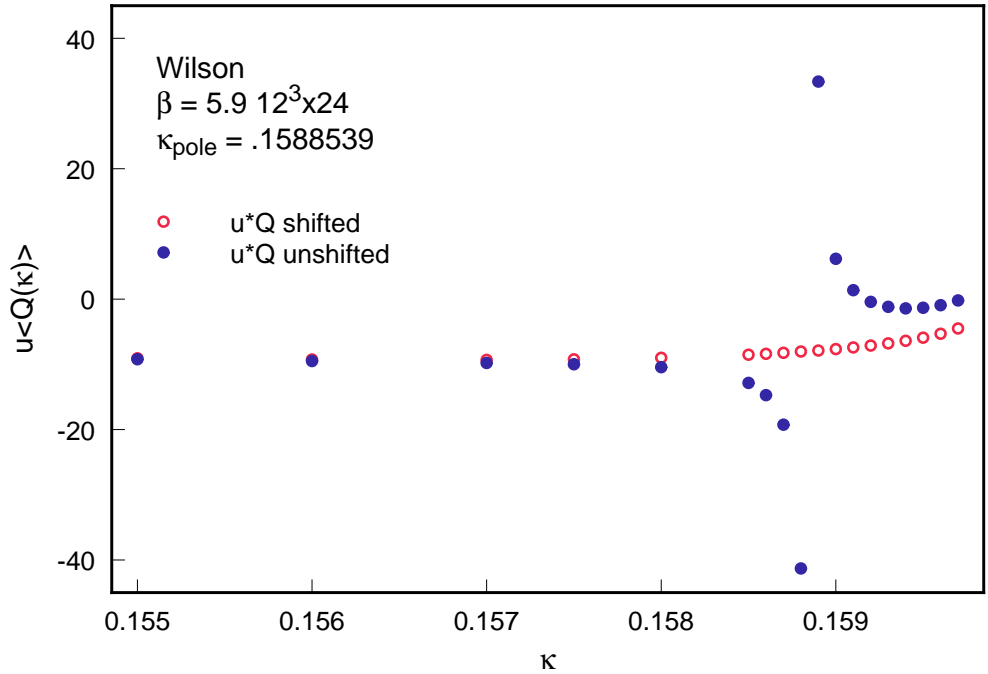}
\caption[]{$u Q(\kappa)$ is shown as a function of kappa for
Wilson action configuration 114000. Here $u$ is twice the 
bare quark mass and $Q(\kappa)$ is the (space-time) average pseudoscalar
charge density. The unshifted charge
(denoted by solid dots) has a pole at $\kappa = .1588539$. 
In contrast the charge in the modified quenched approximation
(denoted by open cirles) shows only smooth behaviour.}
\label{fig:Q5poles}
\end{figure}

In the present calculations we have used eighteen kappa values 
and a Pad\'e method for fitting the pole structure.     This 
procedure works well if the poles are in the visible region, 
$\kappa < \kappa_{max} < \kappa_c$.   Although we had 
originally hoped to determine the complete spectrum of real 
eigenvalues using this method, it was only found to be 
reliable if the poles were in the ``visible" region.   There we 
could determine the pole positions and the relevant 
eigenvalues to great precision.    An example of these scans is 
shown in Figure \ref{fig:Q5poles} for Wilson fermions.
%
%%% begin figure 3
%
%
%%% end figure 3
%
The existence of isolated poles is obvious from this figure. 
The value of the integrated pseudoscalar charge can be computed, 
with no appreciable slowdown in convergence, for values of the hopping parameter
arbitrarily close to the pole position.

For the computations in this paper, we use a sample of 
50 gauge configurations available from the ACPMAPS library, 
the e-lattice set \cite{f_b94}.    These configurations were generated 
on a $12^3\times 24$ lattice at $\beta = 5.9$ and are separated by 
2000 sweeps. We determine pole eigenvalues for
both the standard Wilson-Dirac action and a Clover 
action with a clover coefficient  of  $C_{SW} = 1.91$ 
corresponding to the value suggested by L\"uscher \cite{Luscher96}.   
We find six configurations with visible poles for each choice
of fermion action.    
These results are shown in Table I.    It is important to note 
that only one gauge configuration is in common between the 
two sets of visible poles.    This emphasizes the point that the 
existence of visible poles for a particular gauge configuration 
is a sensitive artifact of the particular choice of fermion action.    
As the fermion action is varied, the real eigenvalues 
move, some to smaller values of kappa where they may 
become visible and some to larger values of kappa where they 
may not be observed as visible poles.    Therefore, the visible 
poles, and the corresponding identification of exceptional 
configurations is a sensitive property of the fermion action 
and not a property unique to the particular gauge 
configuration.    In the same vein, a small change in the 
clover coefficient may remove a visible pole for one 
configuration and add a visible pole for another 
configuration, completely changing identification of the 
exceptional configurations. Since only a collision of two real modes
allows the pairing required to move off the real axis,   
small changes in the parameters of the fermion action should 
not change the number of isolated real modes but only their visibility.

\begin{table}
\begin{center}
\caption{Visible poles for a set of 50 gauge configurations (102000-200000 
step 2000).}
\begin{tabular}{|c|c|c||c|c|c|}
\hline
\multicolumn{3}{|c|}{Wilson Action}&\multicolumn{3}{|c|}{Clover Action, CSW = 1.
91} \\
\hline
Conf. & $\kappa_{pole}$&        PS residue& Conf. &$\kappa_{pole}$&PS residue\\
\hline
114000 &0.1588539&      -2.1463 & 122000&       0.1331519&      +3.3366 \\
132000 &0.1594870&      -2.4800 & 150000&       0.1339044&      +4.8900 \\
148000 &0.1593216&      -3.6325 & 160000&       0.1334519&      -2.0718 \\
160000 &0.1593803&      -2.8494 & 162000&       0.1335898&      -2.5609 \\
182000 &0.1593803&      +2.5036 & 178000&       0.1334681&      -23.4917 \\
194000 &0.1595557&      -5.3055 & 178000&       0.1337479&      +22.7462 \\
&&                              & 198000&       0.1337620&      +1.5828 \\
\hline
\end{tabular}
\end{center}
\end{table}

\begin{figure}
\epsfxsize=14cm
\dofig{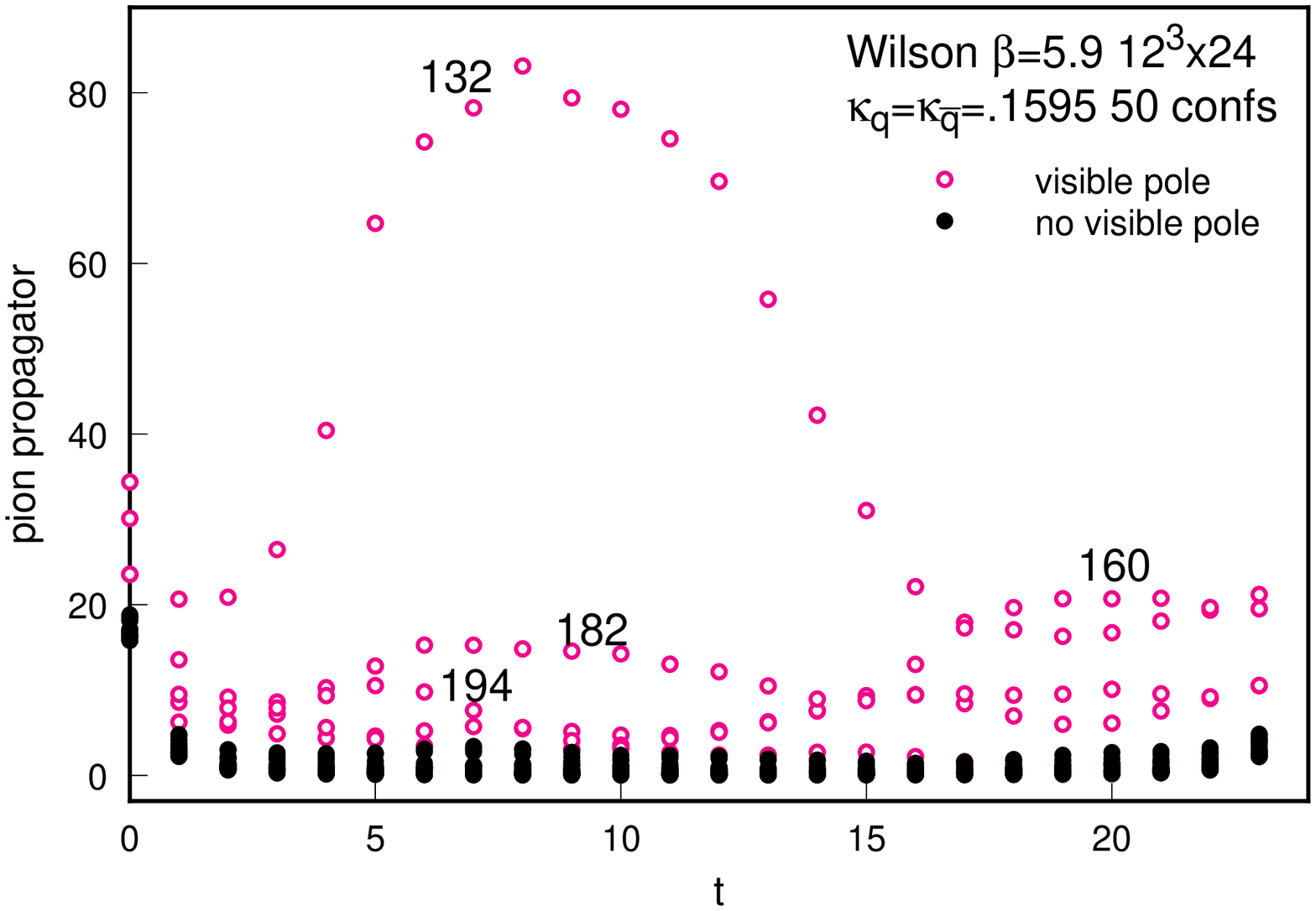}
\caption[]{Pion propagator for $\kappa =.1595 $ 
showing all 50 configurations. 
Propagators for configurations with visible poles 
are displayed with open circles and labeled by their sequence number when
appropriate.}
\label{fig:pionset}
\end{figure}

The connection between the visible poles and the 
exceptional configurations is made clear if we plot the time 
dependence of a pseudoscalar meson propagator 
for a kappa value associated with a light quark, e.g. $\kappa=0.1595$.    
In Figure \ref{fig:pionset}, 
we plot all fifty configurations where the configurations 
% 
%%% begin figure 4
%
%
%%% end figure 4
%
with visible poles have been labeled with open circles.   It is 
clear, that all of the distorted propagators associated with the
exceptional configurations can be uniquely identified 
with the presence of nearby visible poles.

\section{The Modified Quenched Approximation}

In the standard quenched approximation, only the 
valence fermion contributions are calculated explicitly from 
the fermion propagators and the effects of the
fermion loop determinant are ignored.  This 
approximation is very sensitive to singularities of the 
fermion propagator which may be encountered in particular
formulations of lattice fermions.   In this section, we identify 
those singularities which are due to artifacts of  Wilson-Dirac 
fermions in quenched lattice calculations.   We give a 
simple prescription which corrects for the principal artifacts 
encountered when using the Wilson-Dirac fermion
formulation for light quark masses.

As noted in the previous section, the eigenvalue 
spectrum of Wilson-Dirac fermions generally contains a number 
of isolated real eigenvalues.   In the continuum limit, these 
eigenvalues are identified as zero modes and occur at precisely 
zero fermion mass.    In the Wilson-Dirac formulation, these eigenvalues 
shift due to the chiral symmetry breaking generated by 
the Wilson-Dirac action.    Some of these eigenvalues are shifted 
to positive mass which causes singular behavior for the 
fermion propagators computed for a mass near a shifted eigenvalue.    
Because of these shifts, there is no common critical value of 
the fermion mass or hopping parameter, even for a specific gauge 
configuration.  The critical value of the hopping parameter is 
normally defined by the chiral limit as determined by the 
ensemble average of the pseudoscalar meson mass over all 
configurations.   In general, these 
eigenvalue shifts would be averaged to zero in the ensemble 
average and no specific corrections would be needed.   
However, in the quenched approximation, the shifts cause 
poles in the fermion propagators which are not  properly 
averaged.   This effect corresponds to similar situations in 
degenerate perturbation theory where small perturbative 
shifts can cause large effects due to small energy 
denominators.   In this case it is known that it is essential to 
expand around the exact eigenvalues and compensate the 
perturbation expansion with counter-terms in each order.    In 
the present case, we argue that a similar compensation must 
be made when using Wilson-Dirac fermions.   Fortunately, we
know that the correct position for the real poles
is a common critical value of the hopping parameter, $\kappa_c$ 
associated with the massless fermion limit.

We are now able to devise a procedure for correcting the 
fermion propagators for the artifact of the visible shifted poles. 
As before, we write the fermion 
propagator as a sum over the eigenvalues of the Wilson-Dirac 
operator,
\begin{equation}
S(x,y;\{U(A)\})_{AB} = \sum_{eigenvalues} 
\frac{f_{iA}(x;U)\bar{g}_{iB}(y,U)}{\lambda_i +1/2\kappa}
\end{equation}
The shifted real eigenvalues cause poles at particular 
values of the hopping parameter.   The residue of the visible poles 
can be determined by computing the propagator for 
a range of kappa values 
close to the pole position and extracting the residue for the full 
propagator.    It is convenient to have first determined the 
pole position very accurately from a global quantity such as the 
integrated pseudoscalar charge before extracting the 
propagator residue.   
With this residue we can define a modified quenched 
approximation by shifting the visible poles to 
$\kappa=\kappa_c$ and adding terms to 
compensate for this shift when the hopping parameter is not near the 
position of the visible poles.     We are able to apply this 
procedure because it removes a specific lattice artifact, and we 
know the correct continuum position of all real eigenvalues.

As an example we can apply this procedure to the 
integrated pseudoscalar charge.   In Equation (11) we 
determined the visible pole position and residue by studying 
the $\kappa$ dependence of $Q(\kappa)$.    Once this has been 
determined we could define a modified quenched approximation 
by moving the visible pole to kappa critical. The 
naive result would be
\begin{equation}
Q(\kappa) \rightarrow Q(\kappa)
-\frac{R}{(1/\kappa-1/{\kappa_{pole}})}
+\frac{R}{(1/\kappa-1/{\kappa_c})}
\end{equation}
The impact of this modification is shown in Figure \ref{fig:Q5poles} for 
Wilson-Dirac fermions. 
The same behaviour is observed in the case of the SW improved Clover action.    
The shift removes the spurious pole and allows a smooth 
extrapolation to zero fermion mass, $\kappa \rightarrow \kappa_c$.    

The above procedure corrects for the leading effects but may 
distort an ensemble average by only shifting visible poles and 
not compensating for poles beyond the visible range. 
Therefore we chose to define the modified quenched approximation (MQA) 
using a compensated shift of the visible 
poles which moves the visible poles while preserving the 
ensemble averages at large mass.   Introducing  a mass 
parameter, u, relative to a common critical value of the hopping
parameter, $\kappa_c$,
\begin{equation}
u=1/\kappa-1/{\kappa_c}=2{m_f}a
\end{equation}
the visible pole may be replaced as follows
\begin{equation}
\frac{1}{u-u_{pole}} \rightarrow \frac{2}{u} - \frac{1}{(u+u_{pole})}
\end{equation}
At large mass the first two terms in the expansion in 
$1/u$ are not modified and terms linear in the shifts should 
average to zero.   Since we are only able to identify poles with 
positive mass shifts, we have compensated a visible pole with 
one shifted to negative mass.   These negative shifts do not 
generate singularities in the fermion propagators computed 
for positive mass values and are expected to cancel against 
poles with negative shifts generated by other configurations in 
the ensemble.   With this procedure, we do not expect any 
large renormalization of $\kappa_c$ 
due to the shifting procedure.  

The full MQA fermion propagator may be simply computed 
by adding a term to the naive fermion propagator which 
incorporates a compensated shift of the visible poles.   The 
modified propagator is given by 
\begin{equation}
S^{MQA}(x,y;\{U(A)\})  \equiv S(x,y;\{U(A)\}) 
+ a_{pole}(\kappa)*{\rm res}_{pole}(x,y)
\end{equation}
where
\begin{equation}
a_{pole}(\kappa) \equiv -\frac{1}{u-u_{pole}} 
+ \frac{2}{u} - \frac{1}{(u+u_{pole})}
\end{equation}
and
\begin{equation}
{\rm res}_{pole}(x,y) \equiv f_{pole}(x,U)\bar{g}_{pole}(y,U)
= \lim_{u \rightarrow u_{pole}}(u-u_{pole})*S(x,y,\{U\})
\end{equation}
As noted above, the propagator residue can be determined by 
computing the fermion propagator at values arbitrarily close 
to the pole position.  The residue of each pole is extracted by 
calculating the propagator at $u_{pole}-\Delta$ and $u_{pole}+\Delta$
with $\Delta\approx 10^{-5}$ where the pole position, 
$u_{pole}$, has been accurately determined from the 
integrated pseudoscalar charge measurement.    
These calculations generate 
only a modest overhead for the computations as only the 
configurations with visible poles need to be corrected.   
For our sample at $\beta=5.9$ ($12^3\times 24$), 
about 10 to 15\% of the configurations 
need to be corrected for visible poles.   At larger physical volumes 
for fixed $\beta$, a larger percentage of configurations 
should be affected; while at higher $\beta$ for fixed
physical volume, a smaller percentage of configurations 
should be affected.

The MQA propagator may now be used to evaluate any 
correlation functions involving Wilson-Dirac fermions.   We 
have simply removed an obvious lattice artifact from the 
fermion propagator which distorts the usual quenched 
approximation.   It is important to note that the artifact is the appearance of visible 
poles at positive mass, not the existence of small or real 
eigenvalues.    It is only the visibility that we have corrected by 
appealing to the correct behavior in the continuum limit.    
We proceed to discuss some applications of the Modified 
Quenched Approximation in the next section.

\section{Applications}

The MQA quark propagators defined in the 
previous section may be used in place of the usual
quenched propagators to compute any hadron physics 
properties accessible in the quenched approximation.   We 
expect that the suppression of the large fluctuations
normally associated with the exceptional configurations
should greatly reduce the errors associated with the 
propagation of light quarks.   The
most sensitive quantities are those associated with the 
chiral limit.   Prime examples are the pion propagator,
the hairpin calculation for the $\eta'$ mass, meson decay
constants and the light hadron spectrum.

\begin{figure}
\epsfxsize=14cm
\dofig{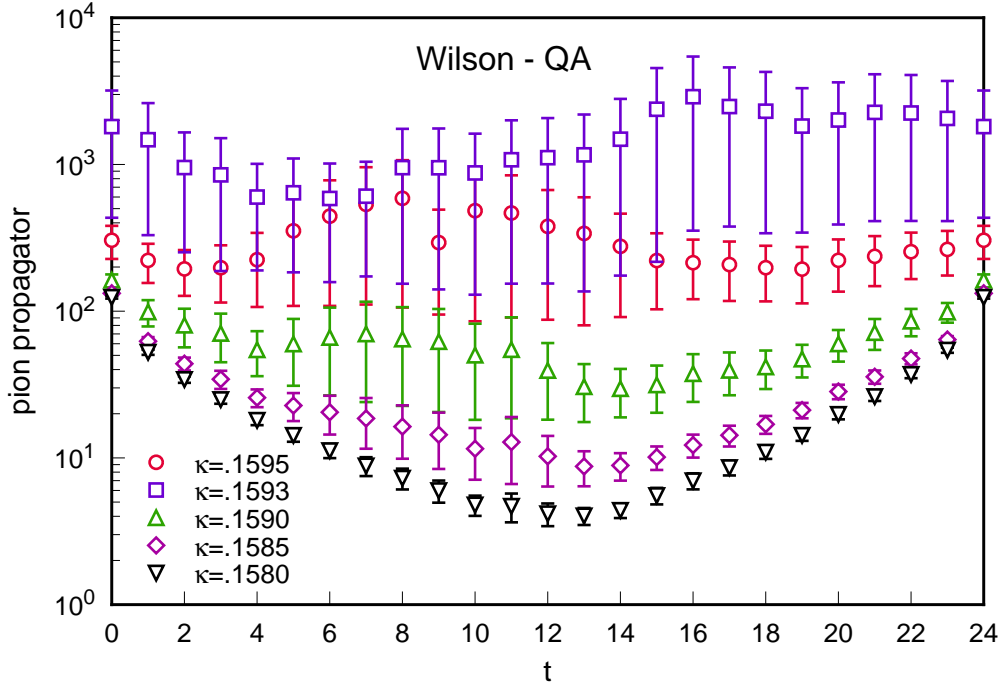}
\caption[]{Pion propagators versus lattice time 
for Wilson action with
naive quark propagators at $\kappa = .1580, .1585, .1590, .1593$
and $.1595$. $m_q = m_{\overline q}$.}
\label{fig:pipropW0}
\end{figure}

\begin{figure}
\epsfxsize=14cm
\dofig{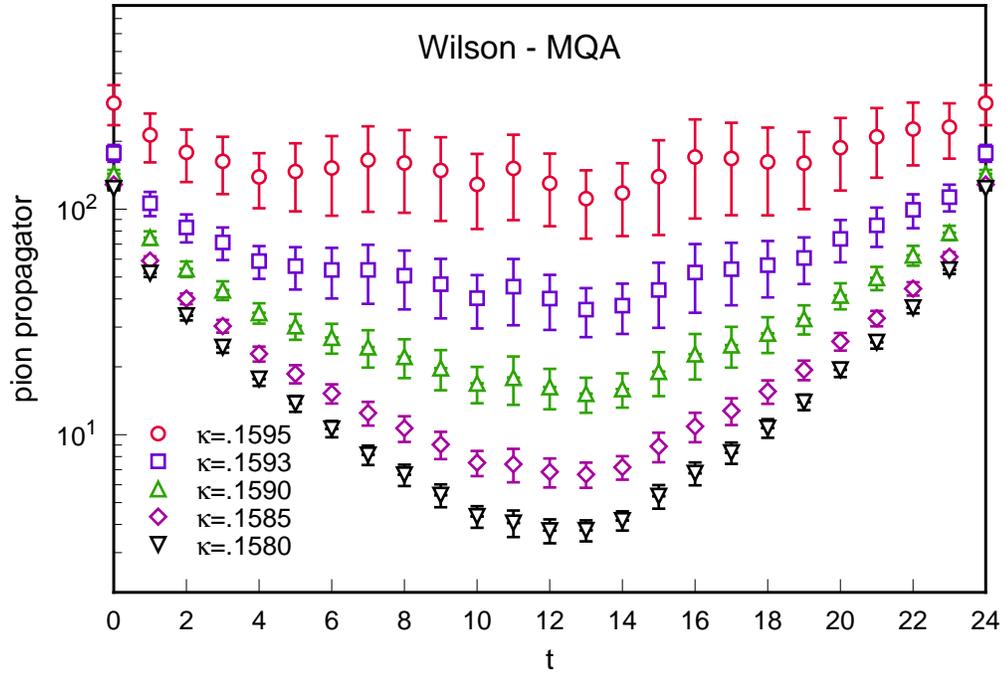}
\caption[]{Pion propagators versus lattice time for Wilson action with
MQA quark propagators at $\kappa = .1580, .1585, .1590, .1593$ 
and $.1595$. $m_q = m_{\overline q}$.}
\label{fig:pipropW1}
\end{figure}

We illustrate the impact of the Modified Quenched
Approximation by computing the pseudoscalar propagator
for a range of quark masses including values appropriate
to physical light quarks.   As mentioned in Section 2, 
we use a sample of 50 gauge configurations generated 
on a $12^3\times 24$ lattice at $\beta = 5.9$, the ACPMAPS
e-lattice set \cite{f_b94}.
We computed the two-point pion correlation
function, $G_{\pi\pi}(t)$, for smeared source and sink 
using an approximation to the pion wavefunction in Coulomb gauge.
In Figure \ref{fig:pipropW0}, we show the time dependence 
of the resulting pion propagators using the usual quark propagators 
obtained for the Wilson-Dirac action for a range of hopping 
parameter values.  
%
%%% begin figure 5
%
%
%%% end figure 5
%
The errors shown are highly correlated.    For 
the larger values of the hopping parameter, corresponding to
the lighter quark masses, the large fluctuations do not 
permit a sensible fit to a pion propagator.
In Figure \ref{fig:pipropW1}, we show the same plot where the quark 
propagators have been replaced by the MQA propagators.
%
%%% begin figure 6
%
%
%%% end figure 6
%
Of the 50 configurations, six  were found to have visible poles with
the pole positions as given in Table I.   The MQA propagators
were computed using the procedures of Section 3 with 
$\kappa_c = 0.15972$.    We find the 
fluctuations are greatly suppressed
allowing a measurement of the pion mass even for the
largest hopping parameter value. (This value of kappa corresponds
to a light quark mass not much larger than the average 
physical up and down masses.)   Even for
heavier masses, the errors appear to be greatly reduced
by using the MQA propagators.

The same analysis has been applied to the case of fermions 
defined through the improved Clover action.   
Again we find that there are
large fluctuations for the hopping parameter values 
corresponding to the light quark limit.    The pion 
propagators computed using the MQA quark propagators 
show greatly reduced fluctuations.    In Table I, there are 
six configurations with visible poles for the Clover action
although one configuration exhibits two visible poles.   Only
one gauge configuration has visible poles for both Wilson-Dirac and
Clover actions.   Again, the MQA analysis is seen to greatly 
reduce the fluctuations of the pion propagator.   Here we have
used $\kappa_c=0.13425$ for the critical hopping parameter.   It 
is clear that clover improvement does not mitigate the problem of
visible poles although the MQA analysis is equally effective
for the Wilson-Dirac and Clover actions.    

\begin{figure}
\epsfxsize=14cm
\dofig{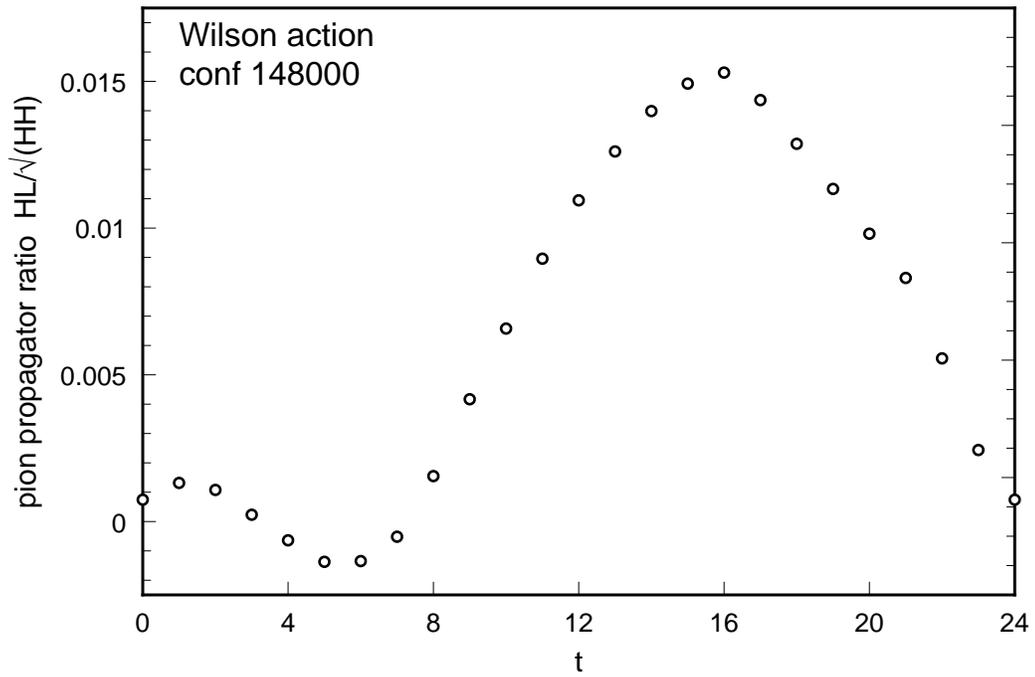}
\caption[]{Real mode pole residue at $\kappa = .1593216$ 
for configuration 148000 with Wilson action.
The residue is computed using the ratio of heavy-light 
pion propagator normalized by the square root of the heavy-heavy propagator.
Here the heavy quark is $\kappa = .1570$ and the light quarks used have
$\kappa = .1595 {\rm -} .1590$.}
\label{fig:resW0}
\end{figure}

The pion propagators we have shown are computed with
the limited statistics of  50 gauge configurations.   
For the lighter quark masses, it is clear that the normal analysis 
would not be limited by statistics but by the frequency of
exceptional configurations associated with 
visible poles.    However, we believe that the MQA analysis
cures this problem and higher statistics would now greatly
improve the accuracy of computations with light mass
quarks.
We can isolate the effect of the real pole by extracting its residue 
from a fit to a series of pion propagators in which only one 
quark mass is varied. The results of that procedure for a particular
configuration (148000) with a pole in the visible region is
shown in Figure \ref{fig:resW0}.
%
%%% begin figure 7
%
%
%%% end figure 7
%

Another way to see the effects of using the MQA analysis 
employs a bootstrap procedure.   We create 200 bootstrap
sets of 50 configurations each randomly
selected from the full set of 50 gauge
configurations.  We simultaneously fit 
the three pion two-point correlators LS(local-smeared), SL(smeared-local)
and SS(smeared-smeared) to a common pion mass.
Each correlator has the form:
\begin{equation}
 G_{\pi\pi}(t) = |A|^2 2 \cosh(m_{\pi}t) \label{eq:fit}
\end{equation}

\begin{figure}
\epsfxsize=14cm
\dofig{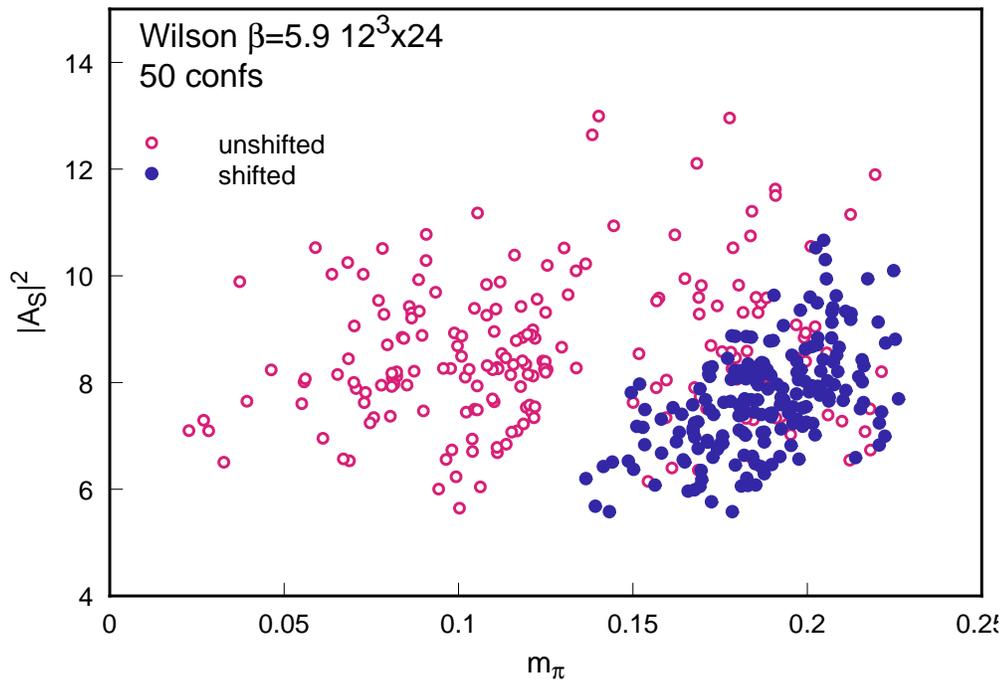}
\caption[]{Pion mass, $m_{\pi}$, and coupling amplitude, 
$|A_S|$, extracted from
the pseudoscalar-pseudoscalar correlator using smeared 
source and sink operators. 
The results for 200 bootstrap samples 
are displayed with naive (open circles) and MQA shifted (solid dots)
Wilson action quark propagators.}
\label{fig:Wboot}
\end{figure}

We determine fits to the pion mass, $m_{\pi}$, and 
coupling amplitudes, $|A_L|$ and $|A_S|$, 
for each of the 200 bootstrap sets for 
particular values of the hopping parameter.   In Figure \ref{fig:Wboot}, 
scatter plots compare the fluctuations and correlations for 
Wilson-Dirac fermions before and after using the MQA
analysis.    
%
%%% begin figure 8
%
%
%%% end figure 8
%
It is clear that the MQA analysis greatly reduces
the fluctuations and produces a more tightly correlated
fit for the mass and decay constant.   Again, we would 
expect even more dramatic  improvement with a larger 
statistics sample of gauge configurations.

\begin{figure}
%\epsfxsize=14cm
%\dofig{pimassW1.eps}
\vspace*{-0.75in}
\hspace*{-0.75in}\psfig{figure=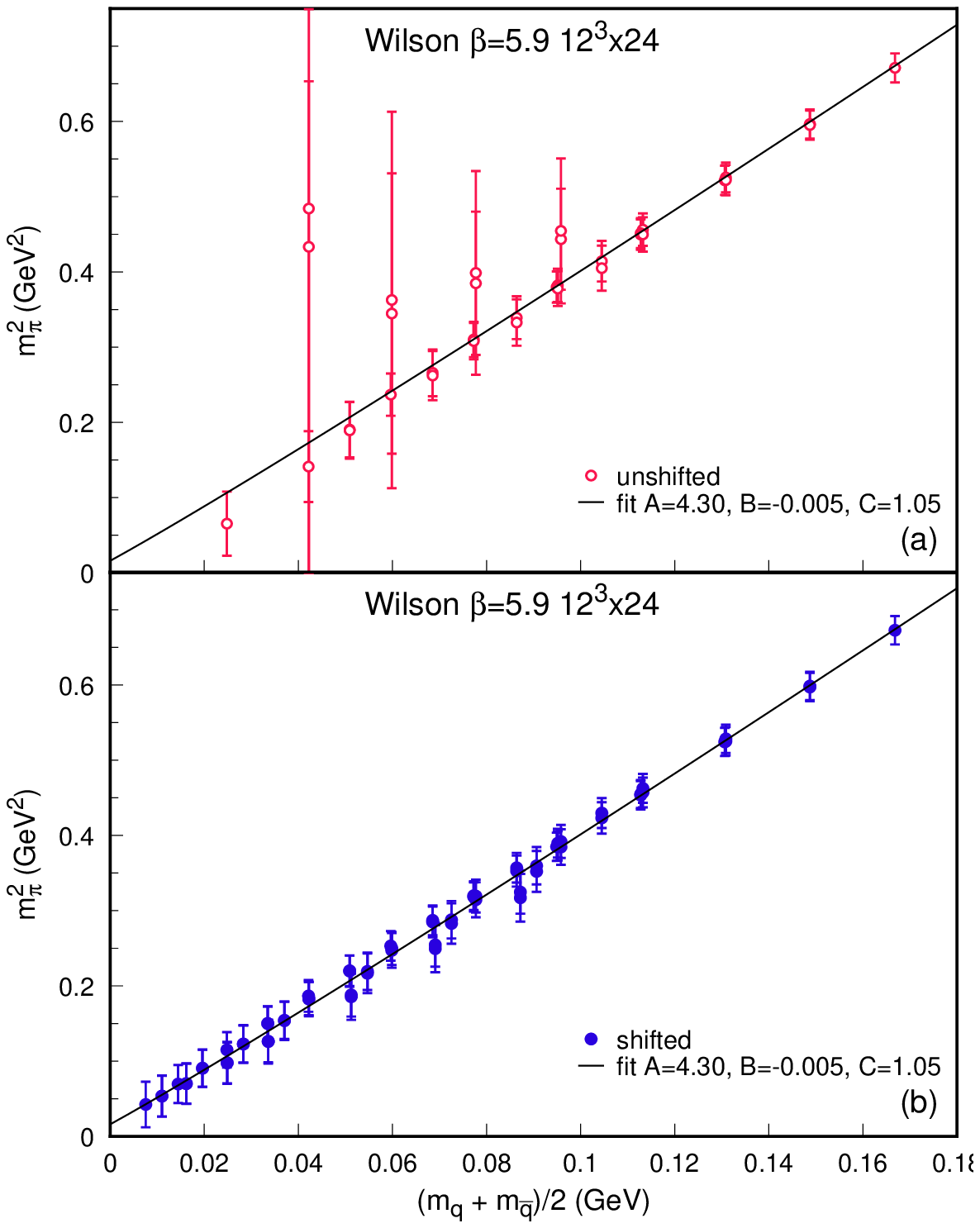}
\vspace*{-4.35in}
\caption[]{Pion mass plots for Wilson action with
(a) naive quark propagators in the kappa range, 
kappa = 0.1550 to 0.1590 (open circles)
and 
(b) MQA propagators in the full kappa range,
kappa = 0.1550 to 0.1595 (dots). 
A power law fit (Eq. 20) to the
MQA masses is also shown.}
\label{fig:pimassW0}
\end{figure}

Using the full bootstrap set, we have made an
uncorrelated fit for the pion mass and decay constants
where the quark masses range over all eight values
of the hopping parameter considered in this paper.   We 
use data on correlation functions for smeared and local 
pseudoscalar sources and the local axial vector charge 
density for our analysis.   
In Figure \ref{fig:pimassW0}, we plot the square
of the measured values of the pion mass for Wilson-Dirac
fermions against the average of the quark masses,
$m_l \equiv (m_{q1}+m_{q2})/2=(0.5/k_{q1} + 0.5/k_{q2} -1/\kappa_c)/(2a)$, 
for the naive and MQA analysis, respectively.    
%
%%% begin figure 9
%
%
%%% end figure 9
%
The large
fluctuations in the naive analysis come when one or both
of the quarks are light.   We show a simple best
fit for the pion mass squared as a function of the average 
bare quark masses, $m_l$, 
assuming the general power law form:
\begin{equation}
  m_{\pi}^2 = A (m_l - B)^C  \label{eq:massfit}
\end{equation}
The masses determined from the 
MQA analysis seem to give a good fit to a nearly linear 
behavior with little evidence for a modified power law (C=1.05).   
The slope determined by our fit is $m_{\pi}^2/m_l = 4.30$
which is in good agreement with previous measurements 
\cite{f_b94}.   The fit also yields a new critical value for
the hopping parameter, $\kappa_c=0.159725$, which should
be compared with the value of $\kappa_c=0.15972$ obtained
from the standard analysis.
The small shift in $\kappa_c$ reflects our use of 
a compensated shift for the visible poles and the linearity
observed in our mass fits.    If uncompensated shifts were
used for the visible poles, or if the value of the critical hopping
parameter were varied slightly, a modest renormalization of
$\kappa_c$ would be expected for the MQA analysis
compared to the normal procedure.    In extreme cases, it may
be necessary to iterate the shift process to determine a 
consistent value of $\kappa_c$.

\begin{figure}
%\epsfxsize=14cm
%\dofig{pimassW1AB.eps}
\vspace*{-0.75in}
\hspace*{-0.75in}\psfig{figure=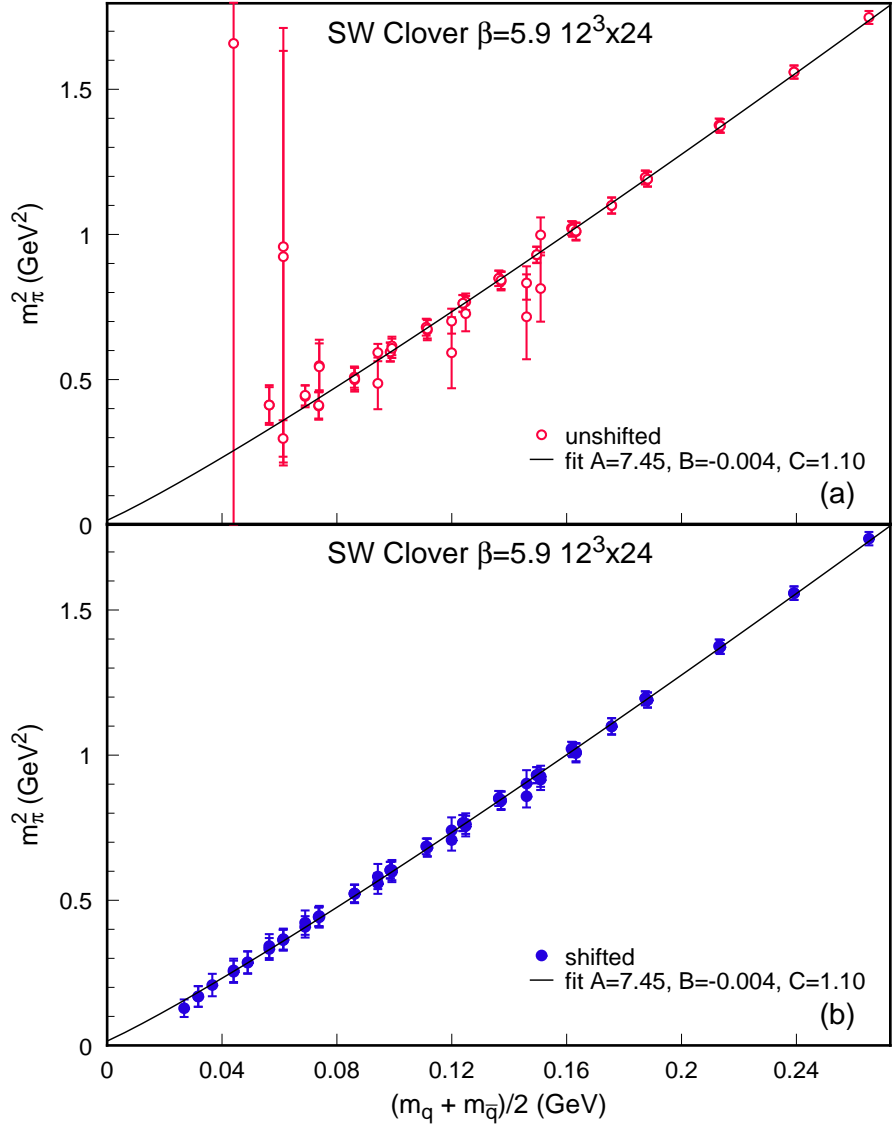}
\vspace*{-4.35in}
\caption[]{Pion mass plots for SW Clover action with
(a) naive quark propagators in the kappa range, 
kappa = 0.1290 to 0.1335 (open circles)
and 
(b) MQA propagators in the full kappa range,
kappa = 0.1290 to 0.1337 (dots).
A power law fit (Eq. 20) to the
MQA masses is also shown.}
\label{fig:pimassW1}
\end{figure}

We also show plots of the fits for
the Clover action with and without MQA analysis in 
Figure \ref{fig:pimassW1}.  
%
%%% begin figure 10
%
%
%%% end figure 10
%
Of course, the slope $m_{\pi}^2/m_l$ here will differ significantly from
the previous case (using naive Wilson action at $\beta = 5.9$).
Using the clover coefficient $C_{SW} =  1.91$,  we find 
$m_{\pi}^2/m_q = 7.45$. Otherwise, the general conclusions are 
the same in the Clover case as in the Wilson case. 

Another quantity with considerable infrared sensitivity is 
the hairpin contribution to the singlet pseudoscalar mass 
term which is thought to be responsible for the large  
$\eta'$ meson mass in quenched QCD.   We follow the
analysis of Kuramashi et al. \cite{Kuramashi94} and use the
all-source propagators discussed previously in computing
the pseudoscalar densities.   The hairpin contribution to
the $\eta'$ propagator can be isolated using appropriate
color projections of the two point correlation function
of the all-source propagators\cite{Xlogs}.   
As in the case of the pion propagator, the large fluctations observed 
for lighter quark masses in the naive analysis are sharply reduced 
by the MQA analysis.
A detailed study of the hairpin propagator with high statistics
is underway and results will be presented elsewhere.

\begin{figure}
\epsfxsize=14cm
\dofig{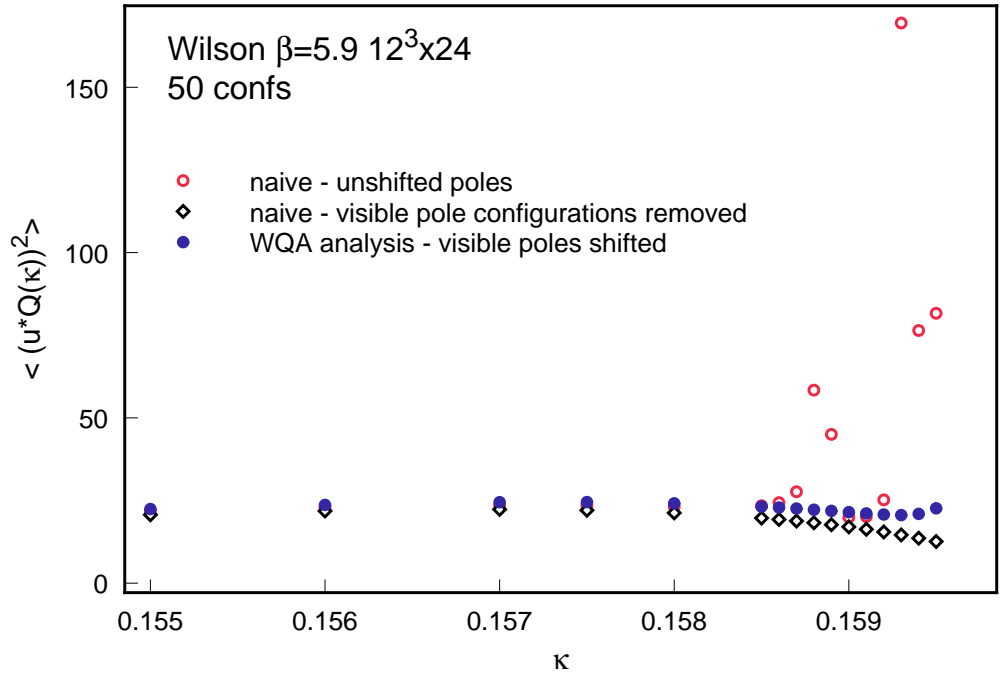}
\caption{$<(u^*Q(\kappa)^2>$ as a function of kappa for Wilson action.
Results are shown for the MQA analysis as well as the naive analysis 
(with and without exceptional configurations).}   
\label{fig:Q5sq}
\end{figure}

In this paper, with our low statistics, we 
restrict our attention to the pseudoscalar charge squared, $Q(\kappa)^2$. 
The weighted average of the square of this charge, 
$<(u^*Q(\kappa))^2>$, as $\kappa$
approaches $\kappa_c$ is the 
average of square of the winding number and therefore a  
measure of the topological susceptibility of QCD. 
In Figure \ref{fig:Q5sq}, we show our
results for the $<(u^*Q(\kappa))^2>$ as a function of kappa the MQA analysis
as well as for the naive analysis (with and without exceptional
configurations excluded).   
%
%%% begin figure 11
%
%
%%% end figure 11
%
As shown in Figure \ref{fig:Q5sq}, the results for naive quenched analysis 
including all configurations is not even smooth as 
$\kappa \rightarrow \kappa_c$.
The full MQA analysis produces a smooth and nearly flat result. 
If we had simply excluded the expectional configurations the result
is smooth but differs in detail from the MQA analysis. In particular,
the value of $\kappa_c$ needed to obtain flat behaviour is shifted.

\section{Discussion and Conclusions}

We have seen that the Wilson-Dirac operator has exact real
modes in its eigenvalue spectrum.   In the quenched approximation,
these real modes can generate unphysical poles in the valence
quark propagators for physical values of the quark mass.    These
poles can produce large lattice artifacts and are the source of
the exceptional configurations observed in attempts to directly
study QCD in the light quark limit.   Indeed, we have observed
poles corresponding to quark masses as large as 20-30 MeV which
is much larger that the physical light quark masses.

We have shown that a simple procedure can be devised to 
completely remove this artifact in realistic calculations.    The
Modified Quenched Approximation identifies and replaces the
visible poles in the quark propagator by the proper zero mode
contribution, compensated to preserve the proper ensemble 
averages at large mass.   The MQA propagator can then be used to compute
all physical quantities normally accessible in the quenched
approximation.    Since only a small number of configurations
require modification (10\% for our lattice at $\beta=5.9$), there
is only a modest overhead cost required to apply the full
MQA analysis.    
As presented here, the only potentially time-consuming part of 
using the MQA analysis is the initial identification of the 
configurations which have visible poles. In this paper, we used
a scan involving the integrated pseudoscalar charge and 18 $\kappa $
values to determine the visible pole positions. 
Subsequently, we have also found that inverting 
only one color-spin component of 
the all source quark propagator and using 12 reasonably spaced 
$\kappa$ values near $\kappa_c$ works well and the computational 
cost is quite modest.
For larger values of $\beta$, a smaller
number of configurations should be affected at fixed physical volume.

The MQA analysis allows stable quenched calculations with
very light quark masses and reduces the errors even in the
case of heavier quark mass.    In the normal analysis, the 
presence of exceptional configurations limits the usefulness of
large statistical samples in many problems.   The
number of exceptional configurations simply grows with
the statistical sample.   With the MQA analysis the 
exceptional configurations are eliminated and the errors
can be meaningfully reduced by using larger statistical
samples.

We have examined both Wilson-Dirac and the improved Clover 
actions.  The Clover action is designed to remove 
O(a) lattice artifacts in the Wilson action.  However, we have found 
that this form of improvement does not remove the problem 
of visible poles and exceptional configurations.   Indeed, we seem 
to find the same size spread of the real eigenvalues for both 
Wilson-Dirac and Clover actions.   A similar conclusion concerning
the frequency of exceptional configurations was reached in a 
different context by L\"uscher \cite{Luscher96}.   

In some applications, the use of improved actions has been 
combined with coarse lattices to avoid the large numerical
overhead associated with very fine grained simulations.   The
MQA analysis may be an essential ingredient in realistic 
applications of these methods.
	
The complete unquenched theory does not suffer 
explicitly from singularities associated with the problem
of shifted poles. The fermion 
determinant is formally the product of the eigenvalues of the 
Wilson-Dirac operator, $(\lambda_i+1/2\kappa)$.   When multiplied by 
the fermion determinant, all of the poles in the sum of the 
valence fermion propagators and hairpin terms are canceled.
There are now 
no singularities associated with the real eigenvalues and their 
shifts are simply averaged when summed over the ensemble 
of configurations.  For this cancellation to occur, it is essential 
that the zeros of the fermion determinant match exactly the 
poles of the fermion propagators and hairpin contributions.   
Without this precise cancellation, the shifted real eigenvalues 
still introduce singular terms.   For example, if the mass used 
in the fermion determinant is not exactly the same as the 
mass used in the propagators, the cancellation fails.   
Hence, it may still be necessary to use the MQA analysis in 
conjunction with an approximate evaluation of the fermion 
determinant in realistic unquenched simulations.
In this case, the sensitive poles are shifted to a common mass 
value and the remaining averages involve insensitive 
contributions or terms which have been effectively moved to the 
numerator.   Here it is essential to realize that the shifted real 
eigenvalues are artifacts of the specific lattice fermion action 
employed in the calculation and are not to be associated with 
real physical effects of the unquenched theory.

The complex eigenvalues of the Wilson-Dirac action are also 
affected by shifts in the real part of the eigenvalues as is clear 
from the way the doubler states are removed from the 
physical spectrum.   Contributions associated with 
eigenvalues with small imaginary parts may also be sensitive 
to shifts in the real part.   However, artifacts associated with 
these shifts may be suppressed by finite volume effects in 
realistic simulations where there are very few eigenvalues with 
small imaginary parts except for the purely real eigenvalues 
we have previously discussed.   In the infinite volume limit, 
it may be necessary to re-examine the precise structure of all 
small eigenvalues of the Wilson-Dirac action.   We suggest that the 
MQA method should suffice to remove the singular terms 
encountered in realistic calculations using Wilson-Dirac
fermions or similar improved actions.   We have not 
examined other formulations of 
lattice fermions, such as Kogut-Susskind fermions, which 
do not suffer directly from lattice artifacts associated with 
the real eigenvalue spectrum.

\newpage

\section{Acknowledgements}
The computations were performed on the ACPMAPS at Fermilab.
The work of W.B., E.E, and G.H. was performed at 
the Fermi National Accelerator Laboratory,
which is operated by Universities Research Association, Inc., 
under contract DE-AC02-76CHO3000. 
The work of A.D. was supported in part by NSF grant 93-22114. 
The work of H.T. was supported in part by the Department of Energy under 
grant DE-AS05-89ER 40518.


\newpage

\begin{thebibliography}{99}

\bibitem{Mutter86}K.-H. Mutter, Ph. De Forcrand, K. Schilling and R. Somer, 
in Brookhaven 1986, Proceedings, Lattice Gauge Theory, '86, pg. 257.

\bibitem{SWaction}B.-Sheikholeslami and R.~Wohlert, Nucl. Phys. {\bf B259}
(1985) 572;  G. Heatlie, G.~Martinelli, C.~Pittori, G.~C.~Rossi
and C.~T.~Sachrajda, Nucl. Phys. {\bf B352} (1992) 266.  

\bibitem{Banks80}T. Banks and A. Casher, Nucl. Phys. {\bf B171}
(1980) 103.

\bibitem{tHooft}G. `t Hooft, Phys.Rev.Lett.{\bf 37}, (1976)8.

\bibitem{AStheorem}M. Atiyah and I. Singer, Ann. of Math. 
{\bf 87} (1968) 485; M. Atiyah and G.
Segal, Ann. of Math. {\bf 87} (1968) 531; 
M. Atiyah and I. Singer, Ann. of Math. {\bf 97} (1972) 119, 139.

\bibitem{Smilga92}H. Leutwyler and S. Smilga, Phys. Rev. D {\bf 46},
(1992) 5607.

\bibitem{Smit87}J. Smit and J. Vink, Nucl. Phys. {\bf B286} (1987)485.

\bibitem{SmitVink87}J. Smit and J. Vink, Nucl. Phys. {\bf B284} 
(1987) 234; J. Vink, Nucl. Phys. {\bf B307} (1988) 549.

\bibitem{QED2d97}W. Bardeen, A. Duncan, E. Eichten and H. Thacker, 
Fermilab-PUB-97/119-T.

\bibitem{Davies88}R. Setoodeh, C.T.H. Davies and L.M. Barbour, 
Phys. Lett. {\bf B213} (1988)195. L.~M.~Barbour et al., 
Phys. Rev. D {\bf 46} (1992) 3618.

\bibitem{Verbaarschot96}J.J.M Verbaarschot, hep-lat/96-06009

\bibitem{Iwasaki89} Y. Iwasaki, Nucl. Phys. B (Proc. Suppl.) {\bf 9}
 (1989)254.

%\bibitem{Shuryak93}E.V. Shuryak and J.J.M. Verbaarschot, 
%Nucl. Phys. {\bf A560} (1993) 306.
%"Random Matrix Theory and Spectral Sum Rules for the Dirac Operator in 
%QCD"

\bibitem{Itoh87}S. Itoh, Y. Iwasaki and T. Yoshie, 
Phys. Lett. {\bf B184} (1987)375; S. Itoh, Y. Iwasaki and T. Yoshie, 
Phys. Rev. D {\bf 36} (1987) 527.

\bibitem{Forcrand88} Ph. De Forcrand, R. Gupta, S. Gusken, K.-H. Mutter, 
A. Patel, K. Schilling and R. Sommer, Phys. Lett. {\bf B200} (1988)
143.

%\bibitem{Kuramashi93}Y. Kuramashi, M. Fukugita, H. Mino, M. Okawa 
%and A. Ukawa, Phys. Rev. Lett. {\bf 71} (1993) 2387.

\bibitem{Kuramashi94}Y. Kuramashi, M. Fukugita, H. Mino, M. Okawa 
and A. Ukawa, Phys. Rev. Lett. {\bf 72} (1994)3448.

\bibitem{Xlogs}A. Duncan, E. Eichten, S. Perrucci and H. Thacker, 
hep-lat/96-08110.

\bibitem{f_b94}A. Duncan, E. Eichten, J. Flynn, B. Hill, G. Hockney 
and H. Thacker, Phys. Rev D {\bf 51} (1995) 5101.

\bibitem{Luscher96}M. Luscher, S. Sint, R. Sommer, P. Weisz and U. Wolff, 
hep-lat/96-09035.

\end{thebibliography}
\end{document}